\documentclass[aps,twocolumn,showpacs,prb,preprintnumbers,amsmath,amssymb]{revtex4}
\newcommand{\lsim} 
 {\ \raise.35ex\hbox{$<$}\kern-0.75em\lower.5ex\hbox{$\sim$}\ }
\newcommand{\gsim}
 {\ \raise.35ex\hbox{$>$}\kern-0.75em\lower.5ex\hbox{$\sim$}\ }

\newcommand{\kets}[1]{|#1\rangle}

\newcommand{\bra}[1]{\left<#1\right|}
\newcommand{\ket}[1]{\left|#1\right>}

\def\journal #1#2#3#4{#1 {\bf #2}, #3 (#4)}

\def\PRB{Phys.\ Rev.\ B}
\def\PRL{Phys.\ Rev.\ Lett.}

\def\JPSJ{J.\ Phys.\ Soc.\ Jpn.}

\def\PSCA{Physica}

\usepackage{graphicx}
\usepackage{dcolumn}
\usepackage{bm}
\usepackage{amsmath}
\usepackage{times}
\usepackage[usenames]{color}

\usepackage{ulem}

\begin{document}
\title{
Spin-orbit coupling in octamers in a spinel sulfide CuIr$_2$S$_4$: 
Competition between spin-singlet and quadrupolar states, and its relevance to remnant paramagnetism}
\author{Joji Nasu and Yukitoshi Motome} 
 \affiliation{Department of Applied Physics, University of Tokyo, Hongo, 7-3-1, Bunkyo, Tokyo 113-8656, Japan}
\date{\today}
\begin{abstract}
We theoretically investigate magnetic properties in the low-temperature phase with the formation of eight-site clusters, octamers, in the spinel compound CuIr$_2$S$_4$. The octamer state was considered to be a spin-singlet state induced by a Peieirls instability through the strong anisotropy of $d$ orbitals, the so-called orbital Peierls state. We reexamine this picture by taking into account the spin-orbit coupling which was ignored in the previous study. 
We derive a low-energy effective model between $j_{\rm eff}=1/2$ quasispins on Ir$^{4+}$ cations in an octamer from the multiorbital Hubbard model with the strong spin-orbit coupling by performing the perturbation expansion from the strong correlation limit. 
The effective Hamiltonian is in the form of the Kitaev-Heisenberg model but with an additional interaction, a symmetric off-diagonal exchange interaction originating from the perturbation process including both $d$-$d$ and $d$-$p$-$d$ hopping. Analyzing the effective Hamiltonian on two sites and the octamer by the exact diagonalization, we find that there is a competition between a spin-singlet state and a quadrupolar state.
The former singlet state is a conventional one, adiabatically connected to the orbital-Peierls state. 
On the other hand, the latter quadrupolar state is stabilized by the additional interaction, which consists of a linear combination of different total spin momenta along the spin quantization axis. 
In the competing region, the model exhibits paramagnetic behavior with a renormalized small effective moment at low temperature. This peculiar remnant paramagnetism is not obtained in the Kitaev-Heisenberg model without the additional interaction. Our results renew the picture of the octamer state and provide a scenario for the intrinsic paramagnetic behavior recently observed in a muon spin rotation experiment [K. M. Kojima {\it et al.}, Phys. Rev. Lett. {\bf 112}, 087203 (2014)].
\end{abstract}



\maketitle



%
%

%




\section{Introduction}

Interplay between charge, spin, and orbital degrees of freedom is a central issue in strongly correlated electron systems.
It brings about fascinating properties, such as the colossal magnetoresistance and multiferroics in manganites~\cite{Tokura1994,Kimura2003} and the superconductivity in iron-based compounds.~\cite{Kamihara2008} 
Spinels are a family of compounds, that provide a playground for such cooperative effects between the multiple degrees of freedom. 
For instance, the cooperative effects result in successive phase transitions associated with magnetic and orbital orders in $A$V$_2$O$_4$ ($A$=Zn, Mg),\cite{Ueda1997,Mamiya1997} 
a helical dimerization in MgTi$_2$O$_4$,\cite{Schmidt2004} and a formation of seven-site clusters (heptamers) in AlV$_2$O$_4$.\cite{Horibe2006} 
Particularly, in these exotic phenomena, the orbital degree of freedom describing the anisotropy of the electronic cloud plays a key role in their magnetic and elastic properties.

\begin{figure}[t]
\begin{center}
\includegraphics[width=\columnwidth,clip]{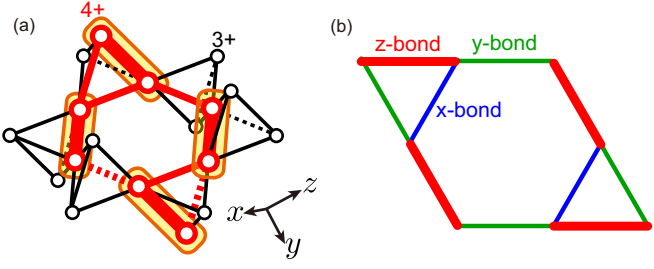}
\caption{(Color online) 
(a) Schematic picture of an Ir$^{4+}$ octamer in the low-temperature phase in CuIr$_2$S$_4$. 
Black thin and red thick open circles represent Ir$^{3+}$ and Ir$^{4+}$ cations, respectively.
The lengths of the shaded bonds are shorter than the others. (b) Schematic picture of the octamer with three kinds of inequivalent bonds. The thick $z$ bonds are shorter than the others.
}
\label{fig:octamer}
\end{center}
\end{figure}

The iridium sulfide CuIr$_2$S$_4$ is one of the spinel compounds in which the multiple degrees of freedom are intricately entangled with each other.~\cite{Nagata1994,Furubayashi1994,Matsuno1997,Nagata1998,Kumagai2000,Radaelli2002,Croft2003,Wang2004,Takubo2005} 
In this compound, the nominal valence of an Ir cation is $3.5+$, which corresponds to the mixed valence state of Ir$^{3+}$ and Ir$^{4+}$. 
This was confirmed by x-ray photoemission spectroscopy~\cite{Matsuno1997} and NMR,~\cite{Kumagai2000} both of which indicate that Cu cations are in the Cu$^+$ state.~\cite{Matsuno1997} The charge degree of freedom is frozen associated with the metal-insulator transition at 230~K. 
The transition is accompanied by the structural change from cubic to tetragonal symmetry.~\cite{Furubayashi1994,Radaelli2002}
In the low-temperature insulating phase, the peculiar charge ordering takes place so that Ir$^{4+}$ cations form eight-site clusters --- octamers, as shown in Fig.~\ref{fig:octamer}(a).
The formation of octamers was elucidated by structural analyses by using synchrotron x-ray diffraction and neutron diffraction; below the charge ordering temperature, it was shown that the system exhibits considerable changes of Ir-Ir bond lengths, leading to the formation of eight-site rings with dimerization in four bonds in each ring, as shown in Fig.~\ref{fig:octamer}(a).~\cite{Radaelli2002}
Moreover, the charge disproportionation of Ir$^{3+}$ and Ir$^{4+}$ was observed by optical conductivity measurement~\cite{Wang2004} and x-ray photoemission spectroscopy.~\cite{Croft2003,Takubo2005} 
The magnetic susceptibility was also measured in CuIr$_2$S$_4$.~\cite{Nagata1998} It shows a sharp drop at the metal-insulator transition from the high-temperature Pauli paramagnetic behavior. In the low-temperature insulating phase, the susceptibility exhibits diamagnetism with less temperature dependence. 
This result suggests the formation of nonmagnetic spin-singlet states in the charge-ordered insulating phase.

A scenario for the metal-insulator transition with the formation of octamers was proposed by considering the orbital degree of freedom in $5d$ electrons in Ir cations.~\cite{Khomskii2005}
In this scenario, the strong anisotropy of $5d$ orbitals under the tetragonal distortion in the low-temperature phase plays a key role as follows.
For the tetragonal distortion which elongates IrS$_6$ octahedra along the $c$ axis, the triply degenerate $t_{2g}$ orbitals of $5d$ electrons split into the higher-energy nondegenerate $xy$ orbital and the lower-energy doubly degenerate $yz$ and $zx$ orbitals. Since there are 5.5 electrons on an Ir$^{3.5+}$ cation, 0.5 holes per site occupy the $xy$ levels. Because of the anisotropy of the $t_{2g}$ orbitals, Ir cations connected by the $\sigma$-bonds between the $xy$ orbitals form one-dimensional chains in the $ab$ planes of the spinel structure. 
Due to the effective quarter filling in the quasi-one-dimensional band, the system is anticipated showing a Peierls instability, which leads to a metal-insulator transition accompanied by dimerization of Ir$^{4+}$ pairs in the $xy$ chains. The resultant fourfold charge ordering in the form of Ir$^{3+}$-Ir$^{3+}$-Ir$^{4+}$-Ir$^{4+}$-$\cdots$ along the chains is compatible with the octamer pattern in the experimental results. Hence, in this scenario, the octamer formation is understood by the Peierls instability induced by the orbital anisotropy, called the orbital Peierls instability.

On the other hand, the spin-orbit coupling in $5d$ electron systems is known to be large in comparison with that in $3d$ and $4d$ electron systems. Recently, it was pointed out that the Mott transition in the layered perovskite Sr$_2$IrO$_4$ is induced not only by the Coulomb interaction but also by the relativistic spin-orbit coupling in $5d$ electrons of Ir cations.~\cite{Kim2008,Kim2009} 
In the Ir$^{4+}$ cation, the strong spin-orbit coupling splits $t_{2g}$ orbitals into $j_{\rm eff}=1/2$ doublet and $j_{\rm eff}=3/2$ quartet, and there is one hole in the $j_{\rm eff}=1/2$ states. The Coulomb interaction may result in the localization of holes in the $j_{\rm eff}=1/2$ narrow band. The resultant insulating state is called the spin-orbit Mott insulator. 

Stimulated by such arguments, effective interactions between the $j_{\rm eff}=1/2$ states in the spin-orbit Mott insulator were theoretically studied for understanding of the low-energy physics. In particular, in the case where octahedra composed by ligands surrounding an Ir$^{4+}$ cation share their edges with each other, the low-energy Hamiltonian includes the peculiar effective ferromagnetic interaction with bond-dependent Ising anisotropy.~\cite{Jackeli2009} 
The same type interaction is found in the Kitaev model, which was exactly shown to be a quantum spin liquid in the ground state.~\cite{Kitaev2006} 
The effective Hamiltonian with the Kitaev interaction in addition to the conventional antiferromagnetic superexchange interaction is called the Kitaev-Heisenberg model. 
This observation has stimulated a hunt for exotic states including a quantum spin liquid in Ir compounds.~\cite{Chaloupka2010,Liu2011,Jiang2011,Reuther2011,Choi2012,Ye2012,Rau2013}

In CuIr$_2$S$_4$, the octamers were suggested to possess Ir$^{4+}$ cations and IrS$_6$ octahedra share their edges in the spinel lattice structure. Therefore, such a Kitaev-type interaction resulting from the strong spin-orbit coupling might also be relevant in this $5d$ electron compound. The effect of the spin-orbit coupling, however, was not taken into account in the orbital Peierls mechanism proposed in Ref.~\onlinecite{Khomskii2005}. Recently, intrinsic paramagnetic behavior with a small effective magnetic moment was observed by a muon spin rotation ($\mu$SR) experiment at low temperature well below the metal-insulator transition.~\cite{Kojima2014} It is difficult to explain this behavior by the orbital Peierls mechanism as there is no active magnetic degree of freedom remaining in the gapped spin-singlet state.
These motivate a reconsideration of the octamer state by explicitly taking into account the strong spin-orbit coupling.

In this paper, we study the effect of the spin-orbit coupling on the magnetic properties in the octamer state in CuIr$_2$S$_4$. Starting from the multiorbital Hubbard model for $t_{2g}$ orbitals with the strong spin-orbit coupling, we consider a low-energy effective model for the $j_{\rm eff}=1/2$ quasispins on Ir$^{4+}$ cations obtained by the perturbation expansion from the strong correlation limit. In this procedure, the spin-orbit coupling is taken into account in the intermediate states of the second-order perturbations.
The effective Hamiltonian includes an additional term to the Kitaev-Heisenberg model, namely, the symmetric off-diagonal exchange interaction, which does not conserve the total magnetic moment along the spin quantization axis.
The equivalent form was recently obtained in Ref.~\onlinecite{Rau2013} for honeycomb-lattice iridium oxides. 
Before considering an octamer, we start with the analysis of the effective Hamiltonian for two spins. We find that the lowest-energy state of the effective model is given by either a spin-singlet state or a quadrupolar state depending on the parameters. 
The former is the conventional singlet state stabilized by the antiferromagnetic Heisenberg interaction. On the other hand, the latter quadrupolar state is the eigenstate of the additional symmetric off-diagonal interaction. The characteristics of this state are that (i) it is described by a linear combination of two-spin states with different total spins along the spin quantization axis and (ii) the coefficient in the linear combination includes a complex phase. We find that these two states compete with each other by changing the spin-orbit coupling, transfer integrals, and Hund's-rule coupling. This competition is also important for the magnetic state in an octamer. Indeed, we find that it brings about peculiar paramagnetic behavior with a small effective moment at low temperature. This intrinsic behavior is not found in the Kitaev-Heisenberg model without the additional term. Our results provide a possible explanation for the remnant paramagnetism recently observed by the $\mu$SR experiment.~\cite{Kojima2014}

This paper is structured as follows. In Sec.~\ref{sec:model}, we present the derivation of the effective Hamiltonian. The multiorbital Hubbard model with the spin-orbit coupling is introduced in Sec.~\ref{sec:multiorbital Hubbard model}. From this Hamiltonian, we derive the low-energy effective model by using perturbation expansion from the strong coupling limit in Sec.~\ref{sec:effective model}. In Sec.~\ref{sec:some remarks}, we remark upon the notable characteristics in the effective Hamiltonian. We also comment on the parameters in the effective model for the dimerization in the octamer in Sec.~\ref{sec:octamer model}.
In Sec.~\ref{sec:method}, we present a numerical method to analyze the effective model. 
In Sec.~\ref{sec:method-result}, we show the results of our numerical analysis.
Before showing the numerical results on the octamer, we present the analysis of the magnetic states in a two-site system in Sec.~\ref{sec:two-site-system}. 
We find the competition between the spin-singlet and quadrupolar states, which leads to peculiar paramagnetic behavior with a renormalized small magnetic moment at low temperature.
The results for the octamer is summarized in Sec.~\ref{sec:eight-site-system}. 
We show that similar competition occurs also in the eight-site cluster. 
We discuss the parameter range and the origin of the remnant paramagnetic behavior of the magnetic susceptibility in detail. The results are compared with those for the Kitaev-Heisenberg model. 
In Sec.~\ref{sec:discussion}, we discuss our results in comparison with the previous theoretical and experimental results. 
Finally, Sec.~\ref{sec:summary} is devoted to a summary.


\section{Model}\label{sec:model}

\subsection{Multiorbital Hubbard model}
\label{sec:multiorbital Hubbard model}

\begin{figure}[t]
\begin{center}
\includegraphics[width=0.9\columnwidth,clip]{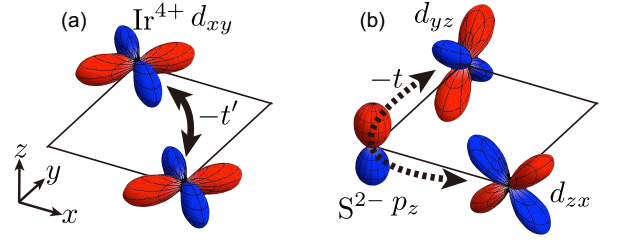}
\caption{
Two types of transfer integrals between Ir$^{4+}$ cations considered in Eq.~(\ref{eq:H_t}): 
(a) $d$-$d$ direct hopping ($-t'$) between $d_{xy}$ orbitals (solid arrow) and 
(b) $d$-$p$-$d$ indirect hopping via $p_z$ orbital between $d_{yz}$ and $d_{zx}$ orbitals (dotted arrow). The transfer integrals for other bonds are obtained by the cyclic permutation of orbital indices.
}
\label{fig:transfer}
\end{center}
\end{figure}

In order to address magnetic properties in the low-temperature octamer phase of CuIr$_2$S$_4$, we start from the following multiorbital Hubbard model for threefold $t_{2g}$ orbitals of Ir $5d$ electrons:
\begin{align}
 {\cal H}={\cal H}_{t}+{\cal H}_{U}+{\cal H}_{\rm SO}.
 \label{eq:H_Hubbard}
\end{align}
The model is defined on the eight-site cluster representing an octamer of Ir$^{4+}$ cations, as shown in Fig.~\ref{fig:octamer}(b). There are three kinds of bonds, $x$, $y$, and $z$ bonds, in this cluster.
The first term ${\cal H}_{t}$ represents the intersite electron transfers; 
\begin{align}
 {\cal H}_{t}=\sum_{
 \langle ij
 \rangle_l}
 (d_{i\gamma\sigma}^\dagger \hat{t}_l^{\gamma\gamma'} d_{j\gamma'\sigma}+ {\rm H.c.}),
 \label{eq:H_t}
\end{align}
where $d_{i\gamma\sigma}$ is the annihilation operator for an Ir hole at site $i$ with orbital $\gamma (=xy,yz,xy)$ and spin $\sigma(=\uparrow,\downarrow)$; the sum is taken for the nearest-neighbor sites $i$ and $j$ on the $l$ bond, and $\hat{t}_l^{\gamma\gamma'}$ is the transfer integral between $\gamma$ and $\gamma'$ orbitals on the $l$ 
bond ($l=x,y,z$) connecting $i$ and $j$ sites.
Here, we take into account two dominant components of the transfer integrals between $5d$ orbitals in the edge-sharing configuration of IrS$_{6}$ octahedra: suppose the bond is a $z$ bond, one is the transfer integral between the $xy$ orbitals coming from a $d$-$d$ direct hopping, and the other is the transfer integral between $yz$ and $zx$ orbitals from a $d$-$p$-$d$ indirect hopping via the $p_z$ orbital at a S$^{2-}$ ion [see Figs.~\ref{fig:transfer}(a) and \ref{fig:transfer}(b), respectively]. 
Namely, we take 
\begin{align}
\label{eq:t}
\hat{t}_z^{xy, xy} &= -t', \\ 
\label{eq:t'}
\hat{t}_z^{yz, zx} &= \hat{t}_z^{zx, yz} = -t, 
\end{align}
and other components zero ($t, t'>0$). 
Here, $t$ is given by $(pd\pi)^2/(\varepsilon_d-\varepsilon_p)$, where $(pd\pi)$ is the Slater-Koster parameter representing the overlap integral between $p_z$ and $d_{yz}$ (or $d_{zx}$) orbitals of S and Ir on the $xy$ plane, and $\varepsilon_d$ and $\varepsilon_p$ are the atomic energies of the Ir $5d$ orbital and  the S $3p$ orbital, respectively. 
For $x$ and $y$ bonds, the transfer integrals are obtained by the cyclic permutation of orbital indices for Eqs.~(\ref{eq:t}) and (\ref{eq:t'}).
We consider the $d$-$p$-$d$ indirect hopping ($-t$) and the $d$-$d$ direct hopping ($-t'$) as free parameters in the following calculations.
We neglect the effect of differences of the bond lengths for a while; it will be incorporated in the model in Sec.~\ref{sec:octamer model}.

The second term in Eq.~(\ref{eq:H_Hubbard}) represents the Coulomb interactions given by
\begin{align}
{\cal H}_{U}&=U\sum_{i\gamma}d_{i\gamma\uparrow}^\dagger d_{i\gamma\uparrow}
d_{i\gamma\downarrow}^\dagger d_{i\gamma\downarrow}
\nonumber \\
&+U'\sum_{i\sigma\sigma'}\sum_{\gamma>\gamma'}d_{i\gamma\sigma}^\dagger d_{i\gamma\sigma}
d_{i\gamma'\sigma'}^\dagger d_{i\gamma'\sigma'}
\nonumber \\
&-J\sum_{i\sigma\sigma'}\sum_{\gamma>\gamma'}d_{i\gamma\sigma}^\dagger d_{i\gamma\sigma'}
d_{i\gamma'\sigma'}^\dagger d_{i\gamma'\sigma} \nonumber \\
&-J'\sum_i \sum_{\gamma>\gamma'} \left (d_{i\gamma\uparrow}^\dagger d_{i\gamma\downarrow}
d_{i\gamma'\uparrow}^\dagger d_{i\gamma'\downarrow} +{\rm H.c.} \right ),\label{eq:1}
\end{align}
where $U$, $U'$, $J$, and $J'$ are the intra-orbital Coulomb repulsion, the inter-orbital Coulomb repulsion, the Hund's-rule coupling, and the pair hopping, respectively. We assume the conditions $U'=U-2J$ and $J'=J$ in Eq.~(\ref{eq:1}).
The last term in Eq.~(\ref{eq:H_Hubbard}) is for the local spin-orbit coupling, which is given by 
\begin{align}
{\cal H}_{\rm SO}=\lambda \sum_i \bm{l}_i\cdot \bm{s}_i.
 \label{eq:H_SO}
\end{align}
Here, $\bm{s}_i$ is the spin of a hole $i$ and $\bm{l}_i$ represents the effective angular momentum for the three $t_{2g}$ orbitals.

\begin{figure}[t]
\begin{center}
\includegraphics[width=\columnwidth,clip]{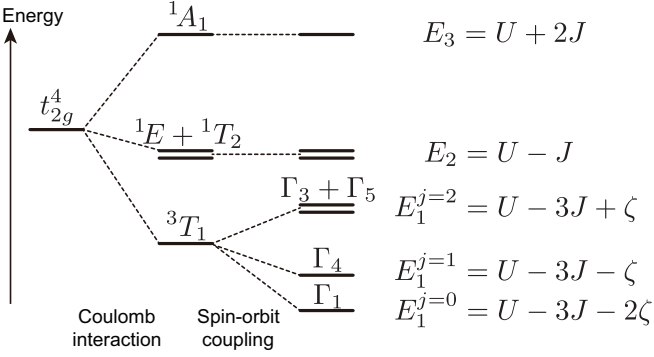}
\caption{
Energy diagram of the $t_{2g}^4$ state by introducing the Coulomb interactions and the spin-orbit coupling. 
The energy for each level is shown in the rightmost.
}
\label{fig:energy}
\end{center}
\end{figure}

\subsection{Effective model in the strong coupling limit}
\label{sec:effective model}

We consider the strong coupling limit of the model in Eq.~(\ref{eq:H_Hubbard}) where the transfer integrals are much smaller than the other energy scales.
When all of the transfer integrals are vanished (${\cal H}_t=0$), the $t_{2g}^5$ state in Ir$^{4+}$ cations splits into $j_{\rm eff}=1/2$ doublet and $j_{\rm eff}=3/2$ quartet by the spin-orbit coupling $\lambda$, and the ground state at each site is given by the $j_{\rm eff}=1/2$ doublet. Note that the $t_{2g}^5$ state is effectively a one-hole state, for which the Coulomb interactions are irrelevant. In order to derive effective superexchange interactions between the $j_{\rm eff}=1/2$ states, we consider the second-order perturbation in terms of the transfer integrals, which induce $t_{2g}^6$ and $t_{2g}^4$ states as the intermediate states. In the independent electron picture, there is 15-fold degeneracy in the $t_{2g}^4$ configuration, while the $t_{2g}^6$ state is nondegenerate. This degeneracy is lifted by the Coulomb interactions and the spin-orbit coupling. Here, we assume the magnitude of the Coulomb interaction is larger than that of the spin-orbit coupling. We note that, for instance, $U$ and $\lambda$ are estimated at about 2~eV and 0.5~eV, respectively, for the iridium oxide Na$_2$IrO$_3$.~\cite{Foyevtsova2013}
First, by introducing the Coulomb interaction ${\cal H}_U$, the 15-fold degeneracy in the $t_{2g}^4$ state is split into four $LS$ multiplets, as shown in the middle of the energy diagram in Fig.~\ref{fig:energy}. The ground state is given by the fourfold ${}^3 T_1$ multiplet. Next, we introduce the spin-orbit coupling ${\cal H}_{\rm SO}$, while neglecting the off-diagonal matrix elements of ${\cal H}_{\rm SO}$ between different $LS$ multiplets. 
Then, the degeneracy of the ground state multiplet ${}^3 T_1$ is lifted as shown in the rightmost of the energy diagram in Fig.~\ref{fig:energy}. Here, the effective spin-orbit coupling for the ${}^3 T_1$ multiplet is written as
\begin{align}
 \tilde{\cal H}_{\rm SO}=-\zeta \sum_i\bm{L}_i\cdot \bm{S}_i,
\end{align}
where $\zeta(>0)$ is given by the reduced matrix element of ${\cal H}_{\rm SO}$ in Eq.~(\ref{eq:H_SO}) for the ${}^3 T_1$ multiplet; $\bm{L}_i$ and $\bm{S}_i$ are the total angular momentum and the total spin moment at site $i$, respectively.

We assume that the magnitudes of the Coulomb interactions ($U,U',J,J'$) and the spin-orbit coupling ($\zeta$) is much larger than those of the transfer integrals ($t,t'$) and that the magnitudes of superexchange interactions ($\sim t^2/U$, $t'^2/U$) are much smaller than the spin-orbit coupling. In this case, $j_{\rm eff}=3/2$ quartet in the $t_{2g}^5$ state can be neglected and the low-energy effective Hamiltonian is given by
\begin{align}
{\cal H}_{\rm eff}={\cal P}_{1/2} {\cal H}_t \frac{1}{{\cal H}_U+\tilde{\cal H}_{\rm SO}} {\cal H}_t {\cal P}_{1/2},
\label{eq:H_eff}
\end{align}
where ${\cal P}_{1/2}$ is the projection operator onto the $j_{\rm eff}=1/2$ doublet. 
After some straightforward calculations, we obtain the effective Hamiltonian on the $\gamma$ bond as follows:
\begin{align}
 {\cal H}_{\rm eff}^{(\gamma)}=J_{\gamma} 
 \sigma_i^\gamma \sigma_j^\gamma
+J_p (\sigma_i^\alpha \sigma_j^\alpha
+\sigma_i^\beta \sigma_j^\beta)
+J_p'(\sigma_i^\alpha \sigma_j^\beta
+\sigma_i^\beta \sigma_j^\alpha),\label{eq:2}
\end{align}
where $(\alpha, \beta, \gamma)$ are the cyclic permutations of $(x,y,z)$. The Pauli matrices $\sigma_i^\gamma$ represent the quasispin operator for the $j_{\rm eff}=1/2$ doublet. The exchange constants are given by
\begin{align}
 J_{\gamma} 
 =&\frac{2t'^2}{27}\frac{1}{E_1^{j=0}}
+\left(
-\frac{t^2}{2}+\frac{t'^2}{18}
\right)\frac{1}{E_1^{j=1}}
\nonumber \\&
+\left(
\frac{t^2}{6}-\frac{t'^2}{54}
\right)\frac{1}{E_1^{j=2}}
+\left(
\frac{t^2}{3}-\frac{t'^2}{27}
\right)\frac{1}{E_2}
+\frac{t'^2}{27}\frac{1}{E_3},
\label{eq:J_z} \\
J_p=&\frac{2t'^2}{27}\frac{1}{E_1^{j=0}}-\frac{t'^2}{9}\frac{1}{E_1^{j=1}}
+\frac{t'^2}{27}\frac{1}{E_1^{j=2}}+\frac{2t'^2}{27}\frac{1}{E_2}
+\frac{t'^2}{27}\frac{1}{E_3},
\label{eq:J_p} \\
J_p'=&\frac{tt'}{3}\frac{1}{E_1^{j=1}}-\frac{tt'}{9}\frac{1}{E_1^{j=2}}
-\frac{2tt'}{9}\frac{1}{E_2}.
\label{eq:3}
\end{align}
Here, $E_1^{j=0}=U-3J-2\zeta$, $E_1^{j=1}=U-3J-\zeta$, $E_1^{j=2}=U-3J+\zeta$, $E_2=U-J$, and $E_3=U+2J$, which are the eigenenergies of the $t_{2g}^4$ states in ${\cal H}_U+\tilde{\cal H}_{\rm SO}$, as shown in Fig.~\ref{fig:energy}. All the exchange processes are taken into account and are characterized by the eigenenergies $E_l$ ($=E_1^{j=0}, E_1^{j=1}, E_1^{j=2}, E_2,$ and $E_3$) of the intermediate states. The coefficients of $1/E_l$ in the exchange constants originate from the transfer processes via the corresponding intermediate state with the eigenenergy $E_l$. Hereafter, we set an energy scale as $t=1$.

\begin{figure}[t]
\begin{center}
\includegraphics[width=\columnwidth,clip]{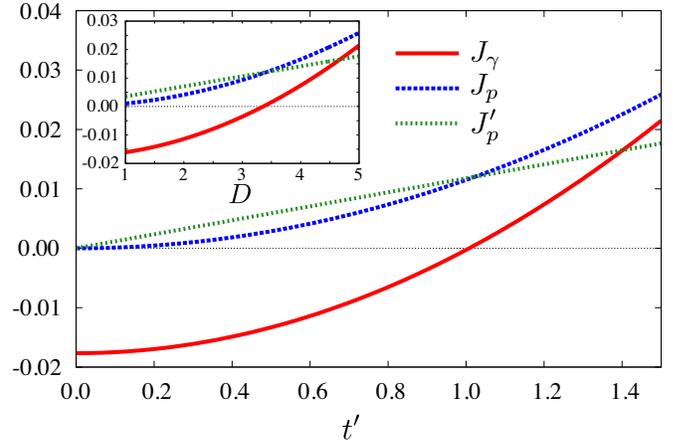}
\caption{(Color online) 
$t'$ dependence of the exchange constants $J_{\gamma}$, $J_p$, and $J_p'$ in Eqs.~(\ref{eq:J_z}), (\ref{eq:J_p}), and (\ref{eq:3}), respectively.
The inset shows the $D$ dependence at $t'=0.3$.
The parameters are chosen to be $U=10$, $J=1$,  and $\zeta=0.5$.
}
\label{fig:exchange}
\end{center}
\end{figure}

\subsection{Some remarks on the effective model}
\label{sec:some remarks}

The effective couplings $J_p$ and $J_p'$ In Eq.~(\ref{eq:2}) are proportional to $t'^2$ and $tt'$, respectively, as shown in Eqs.~(\ref{eq:J_p}) and (\ref{eq:3}). Hence, when $t'=0$, both $J_p$ and $J_p'$ vanish. In this limit, the effective model in Eq.~(\ref{eq:2}) becomes $J_{\gamma} \sigma_i^\gamma \sigma_j^\gamma$ with $J_\gamma <0$, which has the same form as the Kitaev model.~\cite{Kitaev2006,Jackeli2009} It is also noted that when we neglect the $J_p'$ term in Eq.~(\ref{eq:2}), the effective Hamiltonian is the Kitaev-Heisenberg model, which has been studied for several Ir compounds.~\cite{Chaloupka2010,Jiang2011,Reuther2011,Rau2013}
In other words, the effective model in Eq.~(\ref{eq:2}) includes the additional $J_p'$ term to the Kiteav-Heisenberg model.

The $J_p'$ term in Eq.~(\ref{eq:2}) has the form of symmetric off-diagonal exchange interaction. This is derived by the virtual processes where a hole transfers to a neighboring site via the $d$-$d$ direct hopping ($-t'$) and returns to the original site via the $d$-$p$-$d$ indirect hopping ($-t$), and vice versa. 
It is worthy noting that the interaction has the same form as the $xy$ component of the spin quadrupolar operator defined on a bond,~\cite{Shannon2006} 
\begin{align}
Q_{xy} = \sigma_i^x \sigma_j^y + \sigma_i^y \sigma_j^x. 
\label{eq:Qxy}
\end{align}
An equivalent model including the $J_p'$ term was recently discussed for honeycomb-lattice iridium oxides.~\cite{Rau2013}

Among Eqs.~(\ref{eq:J_z})-(\ref{eq:3}), the sign of $J_{\gamma}$ depends on $t$ and $t'$. On the other hand, the exchange constants $J_p$ and $J_p'$ are always positive because these are rewritten as 
\begin{align}
\label{eq:J_p_2}
J_p&=\frac29 \frac{t'^2\zeta^2}{E_1^{j=0}E_1^{j=1}E_1^{j=2}}+\frac{2}{27}\frac{t'^2}{E_2}+\frac{t'^2}{27E_3}, \\
\label{eq:J_p'_2}
J_p'&=\frac29 \frac{tt'}{E_1^{j=1}E_1^{j=2}E_2} [J(E_1^{j=1}+E_1^{j=2})+2\zeta E_2+\zeta^2]. 
\end{align}
Figure~\ref{fig:exchange} shows the $t'$ dependence of the exchange constants. 
As typical parameter values, we here set $U=10$, $J=1$, and $\zeta=0.5$. 
Note that similar values of the parameters were obtained
by the first-principles calculation for Na$_2$IrO$_3$.~\cite{Yamaji2014,Foyevtsova2013} 
At $t'=0$, $J_{\gamma}$ is negative, while $J_p=J_p'=0$. In this case, the present model becomes the Kitaev model with a ferromagnetic Ising-type interaction, as mentioned above. With increasing $t'$, all of the exchange constants increase. 
As shown in Eqs.~(\ref{eq:J_p_2}) and (\ref{eq:J_p'_2}), $J_p$ and $J_p'$ are always positive; $J_p'$ is dominant compared to $J_p$ for $t'/t \ll 1$, as $J_p' \propto t'$ but $J_p \propto t'^2$. 
Meanwhile, $J_{\gamma}$ changes its sign from negative to positive at $t' \sim t$ for the current parameter set. Therefore, for $t'/t\gg 1$, $J_\gamma$, $J_p$, and $J_p'$ are all positive, and the model in Eq.~(\ref{eq:H_eff}) favors an antiferromagnetic configuration for neighboring spins. In particular, in the case with $t=J=\zeta=0$, all of the energies in the intermediate states, $E_l$, become $U$. Then, the effective model is reduced into the Heisenberg model with isotropic exchange interactions; $J_{\gamma}=J_p=t'^2/(9U)$ and $J_p'=0$.

\subsection{Effective model on an octamer}
\label{sec:octamer model}

For an octamer of Ir$^{4+}$ cations, corresponding to the dimerization observed in experiments, we take the length of four $z$ bonds shorter than those of $x$ and $y$ bonds, as shown in Fig.~\ref{fig:octamer}; the lengths of $x$ and $y$ bonds are taken to be uniform for simplicity. 
The difference of the bond lengths is taken into account in the modifications of transfer integrals in Eq.~(\ref{eq:H_t}).
The dimerization shortens the distance between NN Ir cations and reduces the Ir-S-Ir angle, but it is expected not to notably change the distance between NN Ir cation and S anion. Moreover, we expect that the $d$-$p$-$d$ indirect hopping ($-t$) does not strongly depend on the Ir-S-Ir angle.
Relying on these expectations, we take into account the effect of dimerization only on the $d$-$d$ direct hopping $t'$.
Namely, we replace $t'$ by $Dt'$ on the four $z$ bonds, where $D>1$ represents the enhancement factor due to dimerization. 
Note that this replacement modifies the exchange constants in Eqs.~(\ref{eq:J_z})-(\ref{eq:3}) in a different manner.
The exchange constants $J_{\gamma}$, $J_p$, and $J_p'$ as functions of $D$ at $t'=0.3$ are shown in the inset of Fig.~\ref{fig:exchange}.


\section{Method}
\label{sec:method}

We calculate thermodynamic properties as well as the ground-state properties of the effective model in Eq.~(\ref{eq:H_eff}) by using the numerical exact diagonalization. 
We compute the eigenenergies and the diagonal components of the static magnetic susceptibility defined by the canonical spin-spin correlation as 
\begin{align}
 \chi^{\alpha\alpha}=\frac{1}{N}\frac{1}{Z}\sum_{ij}\int_0^\beta d\tau {\rm Tr}[e^{-(\beta-\tau){\cal H}_{\rm eff}}\sigma_i^\alpha e^{- \tau {\cal H}_{\rm eff}}\sigma_j^\alpha],\label{eq:4}
\end{align}
where $\alpha=(x,y,z)$, $\beta=1/T$ is the inverse temperature (we take the Boltzmann constant $k_{\rm B}=1$), $Z={\rm Tr} 
\exp(-\beta{\cal H}_{\rm eff} 
)$ is the partition function, and $N$ is the total number of sites on the cluster. 
In the next section, first, we show the results for a two-site cluster ($N=2$), and then, those for an octamer ($N=8$).


\section{Result}\label{sec:method-result}

\subsection{Two-site system}\label{sec:two-site-system}

\subsubsection{Eigenstates and eigenenergies}
\label{sec:eigenstates_2site}

\begin{figure}[t]
\begin{center}
\includegraphics[width=\columnwidth,clip]{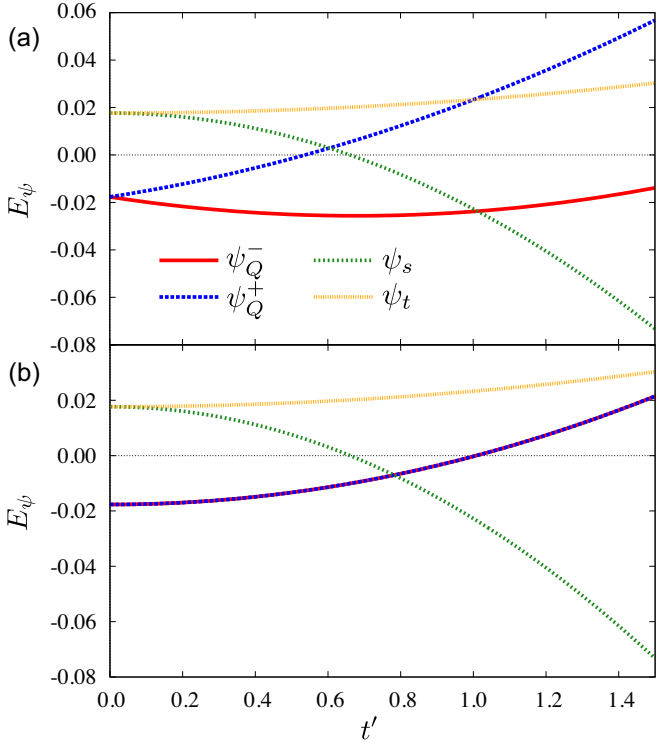}
\caption{(Color online) 
$t'$ dependence of the eigenenergy $E_{\psi}$ in (a) the effective Hamiltonian in Eq.~(\ref{eq:2}) and (b) the model with $J_p'=0$ on a two-site cluster. The eigenstates $\psi_Q^-$, $\psi_Q^+$, $\psi_s$, and $\psi_t$ are given in Eqs.~(\ref{eq:psi_t-})-(\ref{eq:psi_t}). The parameters are chosen to be $U=10$, $J=1$, and $\zeta=0.5$.
}
\label{fig:dimer_ene}
\end{center}
\end{figure}

\begin{figure}[t]
\begin{center}
\includegraphics[width=\columnwidth,clip]{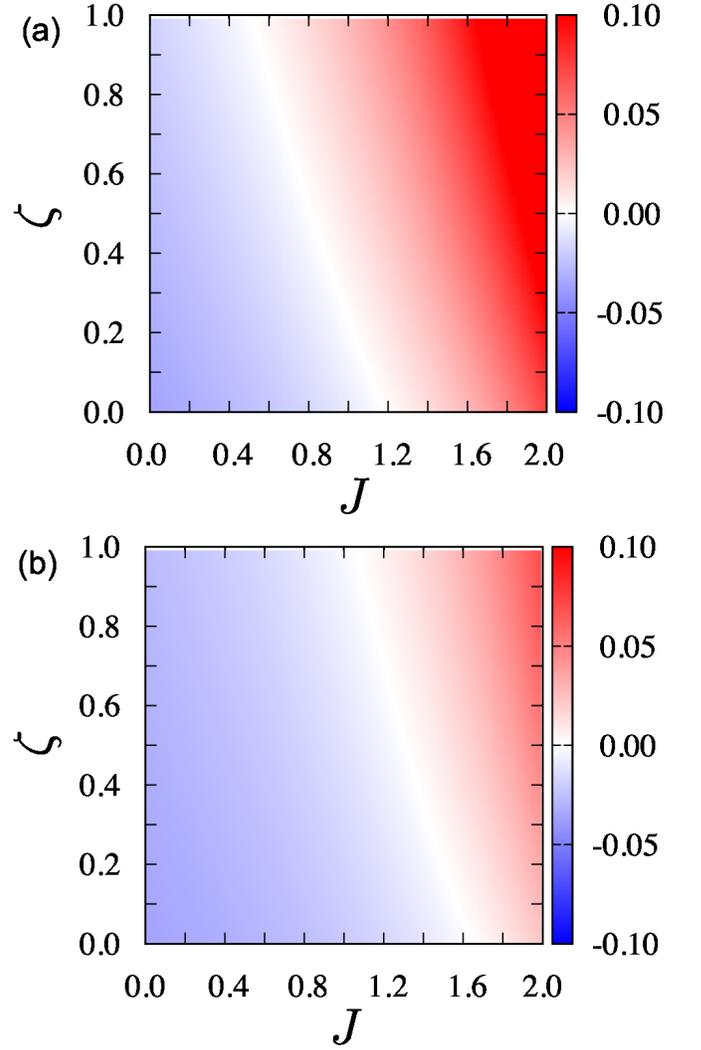}
\caption{
Energy difference between two 
states, $E_{\psi_s}-E_{\psi_Q^{-}}$ 
in (a) the effective Hamiltonian in Eq.~(\ref{eq:2}) and (b) the model with $J_p'=0$ on a two-site cluster. 
The ground state is $\kets{\psi_s}$ ($\kets{\psi_Q^-}$) in the blue (red) region corresponding to the large (small) $J$ and $\zeta$ region. 
In the case with $J_p'=0$, the quadrupolar state $\kets{\psi_Q^-}$, which is the ground state in the red region in (b), is degenerate with the other quadrupolar state $\kets{\psi_Q^+}$. 
The parameters are chosen to be $U=10$ and $t'=0.9$.
}
\label{fig:J_lambda_2site}
\end{center}
\end{figure}

Before showing the results for an octamer, we present the results for a two-site system, as they will be helpful for capturing the essential physics in the dimerized octamer. Here, we focus on the interaction ${\cal H}_{\rm eff}^{(z)}$ for neighboring two spins on the $z$ bond (we take $D=1$). The eigenstates $\kets{\psi}$ and their eigenenergies $E_{\psi}$ of this two-site Hamiltonian are given by
\begin{align}
\label{eq:psi_t-}
 \kets{\psi_Q^-}=\frac{1}{\sqrt{2}}(\ket{\uparrow\uparrow}
-{\rm i}\ket{\downarrow\downarrow}),\ \ \ \ & E_{\psi_Q^-}=J_z-2J_p',\\
\label{eq:psi_t+}
 \kets{\psi_Q^+}=\frac{1}{\sqrt{2}}(\ket{\uparrow\uparrow}
+{\rm i}\ket{\downarrow\downarrow}),\ \ \ \ & E_{\psi_Q^+}=J_z+2J_p',\\
\label{eq:psi_s}
 \kets{\psi_s}=\frac{1}{\sqrt{2}}(\ket{\uparrow\downarrow}
-\ket{\downarrow\uparrow}),\ \ \ \ & E_{\psi_s}=-J_z-2J_p,\\
\label{eq:psi_t}
 \kets{\psi_t}=\frac{1}{\sqrt{2}}(\ket{\uparrow\downarrow}
+\ket{\downarrow\uparrow}),\ \ \ \ & E_{\psi_t}=-J_z+2J_p.
\end{align}
The eigenstates $\kets{\psi_Q^\pm}$ are the mixed states between two states with different total spin moments along the $z$ direction. These states are the eigenstates of the third $J_p'$ term in Eq.~(\ref{eq:2}) which does not commute with $\sigma_i^z+\sigma_j^z$. 
As mentioned in Sec.~\ref{sec:some remarks}, the $J_p'$ term is written by the quadrupolar operator in Eq.~(\ref{eq:Qxy}), and hence, we call $\kets{\psi_Q^\pm}$ the quadrupolar states. 
On the other hand, the states $\kets{\psi_s}$ and $\kets{\psi_t}$ are the conventional spin-singlet state and the spin-triplet state with zero total spin moment in the $z$ direction, respectively.

Figure~\ref{fig:dimer_ene}(a) shows the eigenenergies as functions of $t'$. When $t'$ is large, the ground state is the spin-singlet state $\kets{\psi_s}$. This behavior is consistent with the consideration for the large $t'$ limit in the previous section. On the other hands, when $t'$ becomes small, the ground state is taken over by the quadrupolar state $\kets{\psi_Q^-}$. 
Thus, there is a competition between the spin-singlet state $\kets{\psi_s}$ and the quadrupolar state $\kets{\psi_Q^-}$ in the intermediate $t'$ region. Note that the competition always occurs between these two states because $E_{\psi_Q^-} \leq E_{\psi_Q^+}$ and $E_{\psi_s} \leq E_{\psi_t}$ 
[$E_{\psi_Q^-} - E_{\psi_Q^+} = -4J_p' <0$ and $E_{\psi_s} - E_{\psi_t} = -4J_p <0$; see Eqs.~(\ref{eq:J_p_2}) and (\ref{eq:J_p'_2}), respectively].

In Fig.~\ref{fig:J_lambda_2site}(a), we plot the energy difference between the spin-singlet state and the quadrupolar state, $E_{\psi_s} - E_{\psi_Q^-}$, as a function of the Hund's-rule coupling $J$ and the spin-orbit coupling $\zeta$.
The ground state is either $\kets{\psi_s}$ or $\kets{\psi_Q^-}$ depending on the parameters. 
At $J=\zeta=0$, the ground state is the spin-singlet state $\kets{\psi_s}$ because the Hamiltonian becomes the isotropic Heisenberg model, as discussed in Sec.~\ref{sec:some remarks}. On the other hand, in the large $J$ and/or large $\zeta$ region, the quadrupolar state $\kets{\psi_Q^-}$ becomes the ground state. This result indicates that the Hund's-rule coupling and the spin-orbit coupling stabilizes the quadrupolar state $\kets{\psi_Q^-}$.
There is a level crossing between the two states in the plane of $J$ and $\zeta$, where the first-excitation energy becomes zero. 

Meanwhile, as discussed in the previous section, if we neglect the $J_p'$ term, the effective interaction becomes the Kitaev-Heisenberg form. 
In this case, the ground state for small $t'$ is doubly degenerate between $\kets{\psi_Q^+}$ and $\kets{\psi_Q^-}$, as shown in Fig.~\ref{fig:dimer_ene}(b). 
Figure~\ref{fig:J_lambda_2site}(b) shows the energy difference $E_{\psi_s}-E_{\psi_Q^-}$ in the effective model with $J_p'=0$. 
Although the energy difference $E_{\psi_s}-E_{\psi_Q^{-}}$ behaves qualitatively similar to that in Fig.~\ref{fig:J_lambda_2site}(a), the excitation gap from the ground state is zero in the region of $E_{\psi_Q^-}<E_{\psi_s}$ due to the degeneracy.

\subsubsection{Magnetic susceptibility}
\label{sec:susceptibility_2site}

\begin{figure}[t]
\begin{center}
\includegraphics[width=0.9\columnwidth,clip]{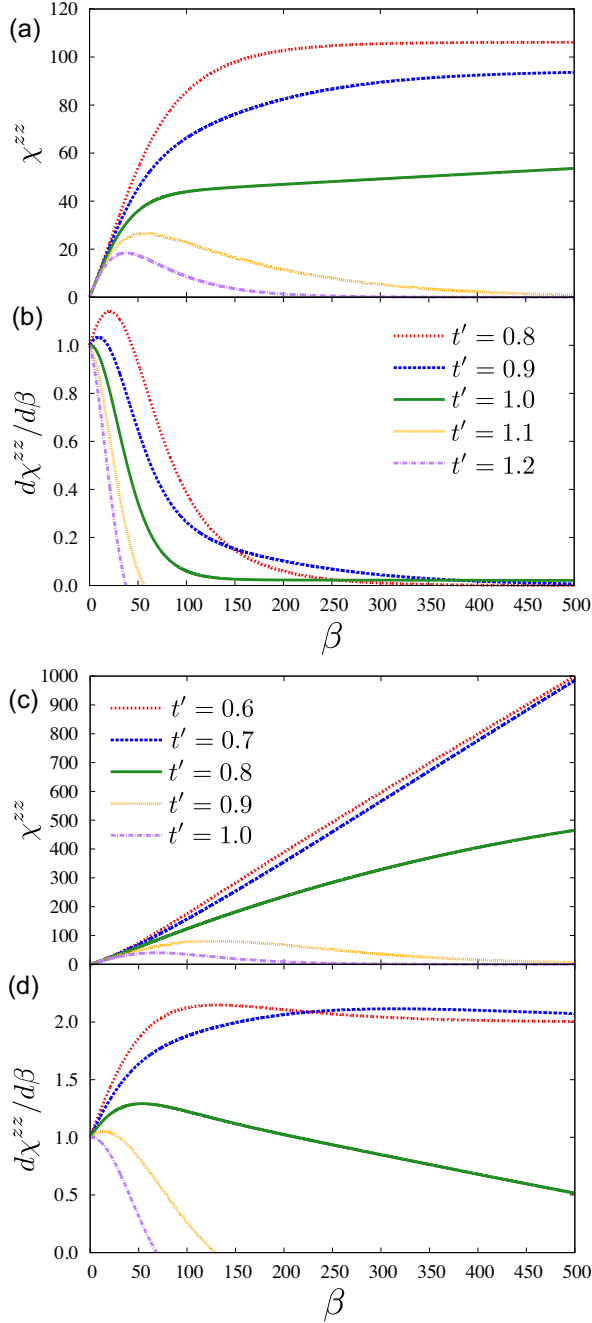}
\caption{(Color online) 
Temperature dependence of (a) the magnetic susceptibility and (b) its $\beta$ derivative for several values of $t'$ in the effective Hamiltonian in Eq.~(\ref{eq:H_eff}) on the two-site cluster.
(c) and (d) show the results when neglecting the $J_p'$ term.
}
\label{fig:sus_2site}
\end{center}
\end{figure}

The competition between the spin-singlet and quadrupolar states gives rise to peculiar temperature dependence of the magnetic susceptibility in Eq.~(\ref{eq:4}). Figure~\ref{fig:sus_2site}(a) shows the calculated data as functions of the inverse temperature $\beta=1/T$ for several $t'$. 
In the high-temperature (small $\beta$) limit, the susceptibility obeys the Curie law and the effective moment given by the slope as a function of $\beta$ is 1 because we choose the magnitude of the local moment at each site as unity (the model is defined by Pauli matrices $\sigma_i^\gamma$ not by $S=1/2$ spins). This is also confirmed by $\beta$ derivative of the susceptibility, as shown in Fig.~\ref{fig:sus_2site}(b). Meanwhile, in the low-temperature (large $\beta$) region, the susceptibility strongly depends on the parameter $t'$. 
When $t'$ is small, the susceptibility increases monotonically with increasing $\beta$ and is saturated at a nonzero value depending on $t'$.
This behavior at low temperature is similar to the Van Vleck paramagnetism because the Hamiltonian does not commute with $\sum_i\sigma_i^z$ and its off-diagonal matrix element between the ground state and excited state is nonzero.

On the other hand, for large $t'$, the susceptibility exhibits a broad peak and turns to decrease down to zero at low temperature. 
This temperature dependence is similar to that in the two-site isotropic Heisenberg model which has the nonmagnetic spin-singlet ground state. The characteristic temperature at which the susceptibility takes a maximum value is determined by the excitation energy. For instance, at $t'=1.2$, the susceptibility becomes maximum at 
$T = \beta^{-1} \simeq 0.02-0.03$, as shown in Fig.~\ref{fig:sus_2site}(a), which corresponds to the energy scale of the gap $E_{\psi_Q^-}-E_{\psi_s}\simeq 0.02$, as shown in Fig.~\ref{fig:dimer_ene}(a). 

The susceptibility shows the peculiar temperature dependence in the intermediate $t'$ region, where the energies of the spin-singlet and quadrupolar states are almost equal.
In this region, the susceptibility is neither saturated at a nonzero value nor suppressed down to zero up to $\beta\sim 500$ with decreasing temperature, as exemplified in the data at $t'=1$ in Fig.~\ref{fig:sus_2site}(a). The low-temperature part gradually increases linearly with $\beta$, and the slope is much smaller than that at high temperature, as shown in Fig.~\ref{fig:sus_2site}(b). 
Moreover, the slope hardly depends on the temperature above $\beta\simeq 200$.
These indicate that the paramagnetic behavior, which is different from the high-temperature limit, appears at very low temperature, and the effective magnetic moment is strongly renormalized from $1$ to a small value. The peculiar behavior originates from the keen competition between $\kets{\psi_s}$ and $\kets{\psi_Q^-}$.

In order to clarify the role of the $J_p'$ term in Eq.~(\ref{eq:2}), we calculate the magnetic susceptibility by omitting it. For large $t'$, the susceptibility behaves similarly to the case in the presence of the $J_p'$ term, as shown in Fig.~\ref{fig:sus_2site}(c); it shows a peak and turns to decrease to zero at low temperature. 
On the other hand, when $t'$ is small, the susceptibility does not saturate at low temperature and the slope becomes larger than $1$ in the low-temperature region, as shown in Fig.~\ref{fig:sus_2site}(c). 
This is in sharp contrast to the result for the model including the $J_p'$ term shown in Fig.~\ref{fig:sus_2site}(a).
Moreover, as shown in Fig.~\ref{fig:sus_2site}(d), the slope in $\beta$, which corresponds to the square of the effective moment per site, approaches $2$ at low temperature. This result is understood as follows. As shown in Fig.~\ref{fig:dimer_ene}(b), the ground states are doubly degenerate between $\kets{\psi_Q^{\pm}}$. 
Since the doublet states are both written in terms of $\ket{\uparrow\uparrow}$ and $\ket{\downarrow\downarrow}$, the effective moment of the ground states is $2$. This results in the slope of $2^2/N=2$, where $N=2$.

In the intermediate $t'$ region where the three states except for $\kets{\psi_t}$ are degenerate [see Fig.~\ref{fig:dimer_ene}(b)], the result is also distinct from that in the presence the $J_p'$ term. 
The temperature dependence of the susceptibility at $t'=0.8$ is presented in Fig.~\ref{fig:sus_2site}(c). The slope is not strongly suppressed in the low-temperature region, but changes gradually at a function of temperature, as shown in Fig.~\ref{fig:sus_2site}(d). The results indicate that the $J_p'$ term in Eq.~(\ref{eq:2}) plays a crucial role in the paramagnetic behavior with a renormalized effective moment at low temperature.

\subsection{Eight-site system}\label{sec:eight-site-system}

\subsubsection{Energy gap}
\label{sec:gap_octamer}

\begin{figure}[t]
\begin{center}
\includegraphics[width=0.9\columnwidth,clip]{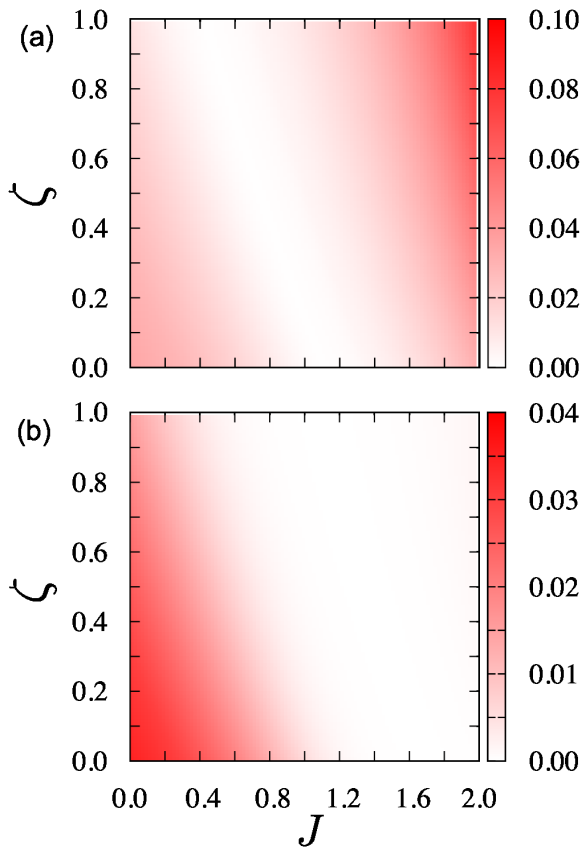}
\caption{(Color online) 
Energy gap between the ground state and the first excited state in (a) the effective Hamiltonian in Eq.~(\ref{eq:H_eff}) and (b) the model with $J_p'=0$ on the eight-site cluster. The parameters are chosen to be $U=10$, $t'=0.3$, and $D=3$.
}
\label{fig:J_lambda}
\end{center}
\end{figure}

In the previous section, we present the results for the two-site system. Here, we analyze the eight-site system which is defined on the octamer shown in Fig.~\ref{fig:octamer}. Figure~\ref{fig:J_lambda}(a) show the energy gap between the ground state and first excited state as a function of the Hund's-rule coupling $J$ and the spin-orbit coupling $\zeta$. In this figure, the parameters are chosen to be $U=10$, $t'=0.3$, and $D=3$. 
Note that the effective interactions on the $z$ bonds for these parameters correspond to those in the two-site system at $t'=0.9$ taken in Fig.~\ref{fig:J_lambda_2site}. The result in Fig.~\ref{fig:J_lambda}(a) is similar to the absolute value of the data plotted in Fig.~\ref{fig:J_lambda_2site}.
This indicates that the energy gap in the octamer is dominated by that in dimers on the $z$ bonds which have a shorter bond length (incorporated by a larger $D$ than $1$). 
Therefore, in the small $J$ and small $\zeta$ region in Fig.~\ref{fig:J_lambda}(a), the ground state of the octamer is approximately described by a direct product of four spin-singlet states $\kets{\psi_{s}}$ [Eq.~(\ref{eq:psi_s})] on the $z$-bond dimers. On the other hand, in the large $J$ and/or large $\zeta$ region, the ground state is close to a direct product of four quadrupolar states $\kets{\psi_Q^-}$ [Eq.~(\ref{eq:psi_t-})] on the $z$ bonds.

We also calculate the energy gap in the Kitaev-Heisenberg model on the octamer by setting $J_p'=0$ in Eq.~(\ref{eq:2}). Figure~\ref{fig:J_lambda}(b) shows the result for the same parameters as those in Fig.~\ref{fig:J_lambda}(a). In the region where both $J$ and $\zeta$ are small, there is a finite gap similar to Fig.~\ref{fig:J_lambda}(a). Hence, in this region, the ground state is considered to be well approximated by a direct product state of $\kets{\psi_s}$ on the $z$ bonds. 
On the other hand, in the large $J$ and/or large $\zeta$ region, the energy gap is vanishingly small, in contrast to the result in Fig.~\ref{fig:J_lambda}(a). This behavior presumably originates from the degeneracy of $\kets{\psi_Q^+}$ and $\kets{\psi_Q^-}$ found in the two-site system.

\subsubsection{Magnetic susceptibility}
\label{sec:susceptibility_octamer}

\begin{figure*}[t]
\begin{center}
\includegraphics[width=1.7\columnwidth,clip]{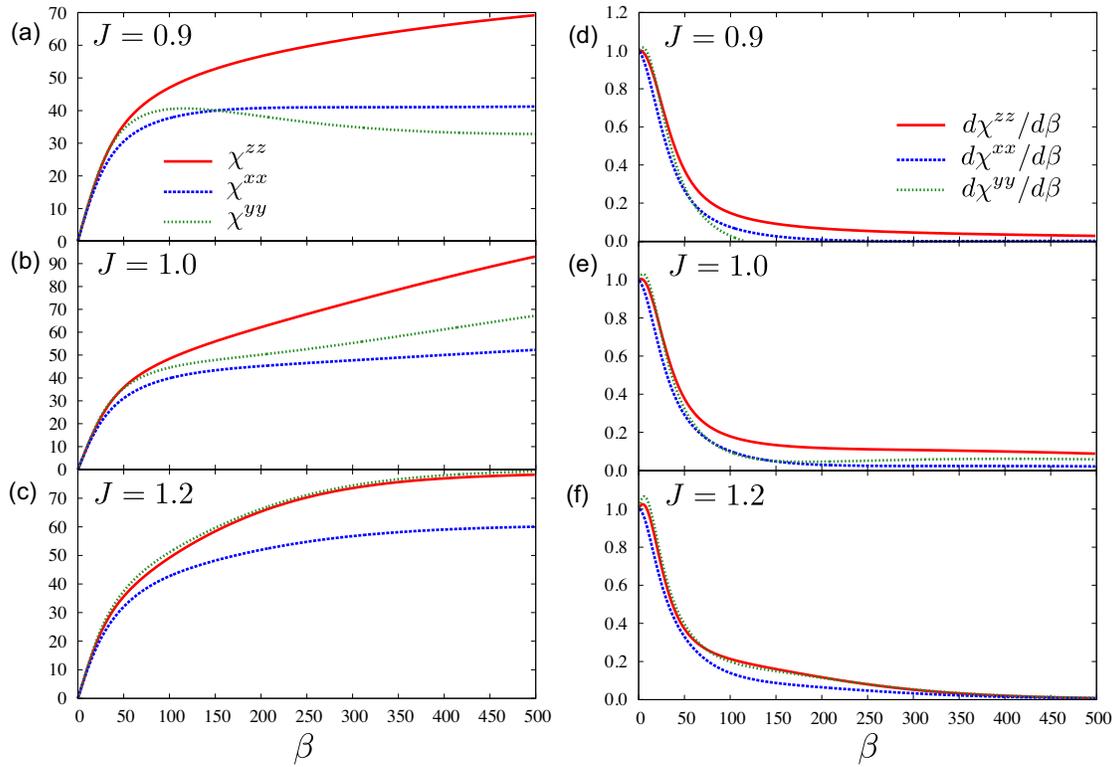}
\caption{(Color online) 
Temperature dependence of (a)-(c) the diagonal components of the magnetic susceptibility and (d)-(f) their derivatives in terms of $\beta$ calculated for the effective Hamiltonian on the eight-site cluster at $J=0.9$, $1.0$, and $1.2$. The parameters are chosen to be $U=10$, $\zeta=0.5$, $t'=0.3$, and $D=3$.
}
\label{fig:sus_full}
\end{center}
\end{figure*}

\begin{figure}[t]
\begin{center}
\includegraphics[width=\columnwidth,clip]{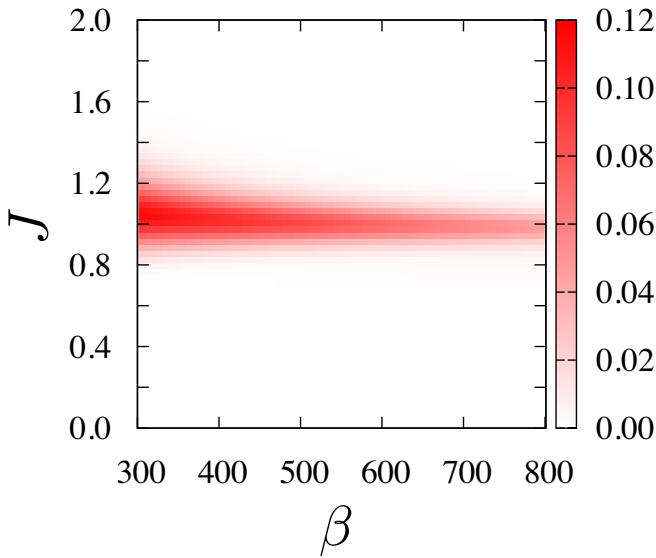}
\caption{(Color online) 
Temperature derivative of the susceptibility $\chi^{zz}$ on a plane of $\beta$-$J$ in the effective model in Eq.~(\ref{eq:H_eff}).
}
\label{fig:contour}
\end{center}
\end{figure}

\begin{figure*}[t]
\begin{center}
\includegraphics[width=1.7\columnwidth,clip]{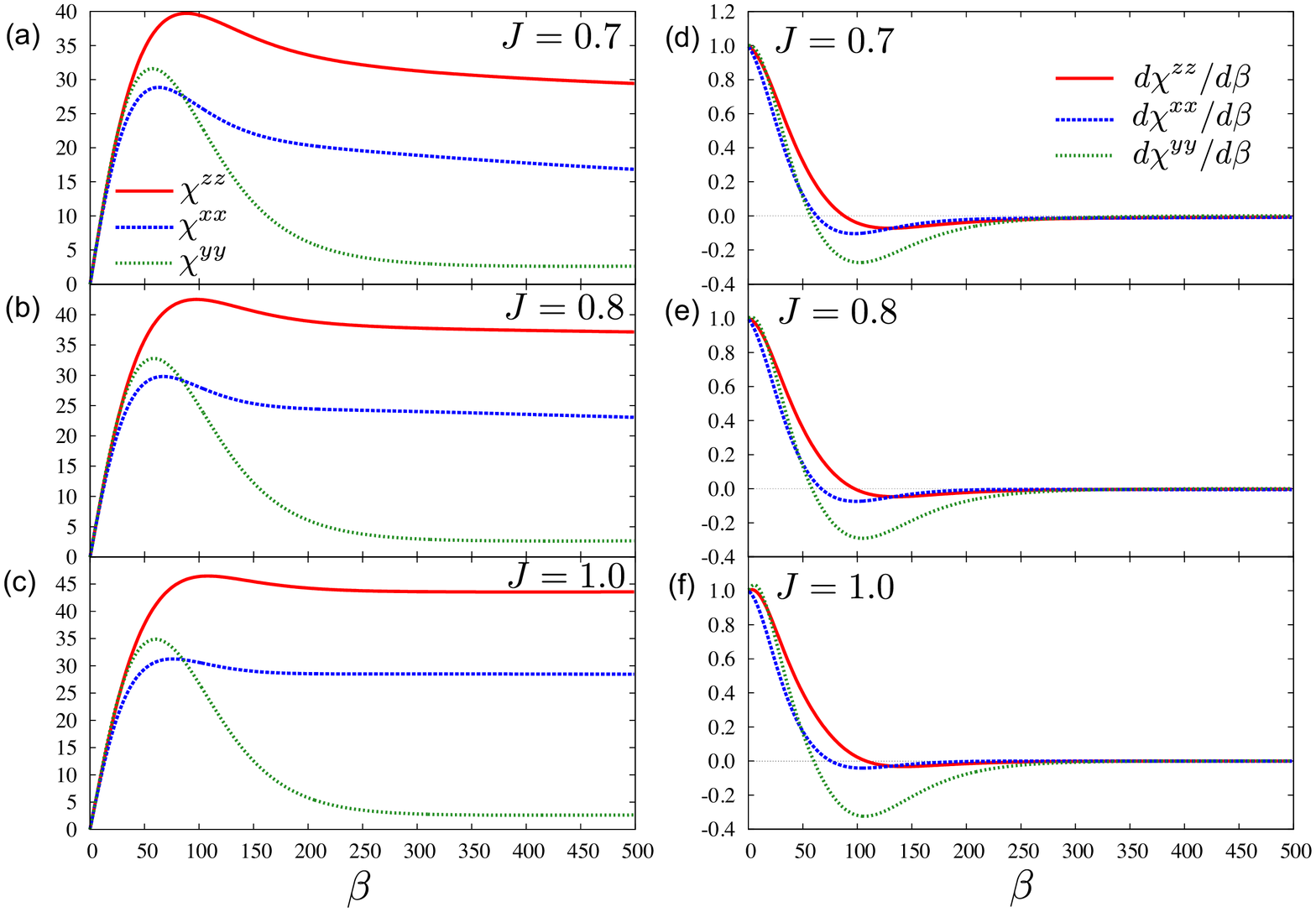}
\caption{(Color online) 
Temperature dependence of (a)-(c) diagonal components of the magnetic susceptibility and (d)-(f) their derivatives in terms of $\beta$ calculated for the effective Hamiltonian with $J_p'=0$ on the eight-site cluster at $J=0.7$, $0.8$, and $1.0$. The parameters are chosen to be $U=10$, $\zeta=0.5$, $t'=0.3$, and $D=3$.
}
\label{fig:sus_KH}
\end{center}
\end{figure*}

Next, we show the temperature dependence of the susceptibility in Fig.~\ref{fig:sus_full} for several values of the Hund's-rule coupling $J$. The data are calculated in the parameter region where the energy gap is small [see Fig.~\ref{fig:J_lambda}(a)]. 
First, let us discuss the results at $J=1.0$ where the energy gap is almost vanishing. Figure~\ref{fig:sus_full}(b) shows the temperature dependence of the susceptibility at $J=1.0$. In this case, all three diagonal components of the susceptibility, $\chi^{xx}$, $\chi^{yy}$, and $\chi^{zz}$, approximately obey the Curie law up to $\beta=500$, and the slopes (effective moments) for $\beta\gtrsim 200$ are substantially smaller than the value of $1$ in the limit of $\beta \to 0$. See also $\beta$ derivatives presented in Fig.~\ref{fig:sus_full}(e);
all three components become $\lesssim 0.1$ and less dependent on $\beta$ for $\beta\gtrsim 200$.
These results indicate that the octamer exhibits peculiar paramagnetic behavior at low temperature and the effective moment is strongly renormalized to a small value at low temperature from the bare moment.

We also calculate the susceptibility while changing $J$ in the vicinity of $J=1.0$.
Figures~\ref{fig:sus_full}(a) and \ref{fig:sus_full}(c) show the temperature dependence of the susceptibility at $J=0.9$ and $J=1.2$, respectively. At $J=0.9$, $\chi^{zz}$ increases with decreasing temperature. This temperature dependence indicates that the renormalized paramagnetic behavior remains apart from $J=1.0$. As shown in Fig.~\ref{fig:sus_full}(d), the slope of $\chi^{zz}$ at low temperature for $J=0.9$ is smaller than that for $J=1.0$. 
For $J \lesssim 0.9$, such remnant paramagnetic behavior at low temperature becomes less distinguished; the susceptibility tends to decrease to zero, similar to the two-site case with large $t'$ in Fig.~\ref{fig:sus_2site}(a).
On the other hand, as shown in Fig.~\ref{fig:sus_full}(c), the susceptibility at $J=1.2$ is not proportional to $\beta$ at low temperature. The slope decreases gradually and becomes almost zero at $\beta=500$, as shown in Fig.~\ref{fig:sus_full}(f). For larger $J \gtrsim 1.2$, the susceptibility tends to saturate at a nonzero value at low temperature, similar to the Van Vleck-type behavior seen for the two-site case with small $t'$ in Fig.~\ref{fig:sus_2site}(a).

Thus, the paramagnetic behavior with a small effective moment at low temperature appears in the parameter region where the excitation gap becomes small. From the argument in Sec.~\ref{sec:gap_octamer}, the small excitation gap is due to the competition between the two different types of dimerized states: one is the state where the four dimers are approximately described by the spin-singlet states $\kets{\psi_s}$, and the other is the state where they are close to the quadrupolar states $\kets{\psi_Q^-}$. The former singlet is stabilized by the antiferro-type Heisenberg interaction which is dominant in the small $J$ and small $\zeta$ region. Meanwhile, the latter quadrupolar state is stabilized by the $J_p'$ interaction which is dominant for large $J$ and/or large $\zeta$.

In order to clarify the parameter region where the paramagnetic behavior with a small effective moment is evident, we show the $\beta$ derivative of the susceptibility on the plane of the inverse temperature $\beta$ and Hund's-rule coupling $J$ in Fig.~\ref{fig:contour}(a). This value takes a nonzero value about or less than $0.1$ not only at $J\simeq 1.0$ but also in between $J\simeq 0.9$ and $J\simeq 1.1$. Furthermore, in this region of $J$, the small but a nonzero value remains by decreasing temperature down to $T\sim 1/800$. 
The effective moment corresponding to the square root of the $\beta$ derivative of the susceptibility takes the highest value $(\sim \sqrt{0.1}\simeq 0.316)$ at $J\simeq 1$, where the energy of the singlet state and the quadrupolar state on dimers compete with each other. When the parameters are apart from the competing point of these two states, the effective moment decreases rapidly.

As shown in the two-site case, the $J_p'$ term plays an important role in the emergence of the remnant paramagnetic behavior.
Figures~\ref{fig:sus_KH}(a)-\ref{fig:sus_KH}(c) show the temperature dependence of the susceptibility at several $J$ calculated by setting $J_p'=0$. 
Since the excitation gap decreases with increasing $J$ and becomes almost zero at $J\simeq 0.8$ for $\zeta=0.5$ as shown in Fig.~\ref{fig:J_lambda}(b), the peculiar temperature dependence is expected to appear at $J\simeq 0.8$ if it exists. The results in Figs.~\ref{fig:sus_KH}(a)-\ref{fig:sus_KH}(c), however, indicate that the susceptibility behaves similarly for different $J$ in this region; 
it obeys the Curie law at high temperature and shows the Van Vleck-type behavior at low temperature.
Indeed, as shown in Figs.~\ref{fig:sus_KH}(d)-\ref{fig:sus_KH}(f), $\beta$ derivatives of the susceptibility approach zero at low temperature. 
Therefore, the paramagnetic behavior with a small effective moment at low temperature is absent in the competing regime when the $J_p'$ term is neglected. This result indicate that the $J_p'$ term is requisite for the paramagnetic behavior at low temperature.

\subsubsection{Spectral decomposition of the magnetic susceptibility}
\label{sec:spectral_octamer}

\begin{figure}[t]
\begin{center}
\includegraphics[width=\columnwidth,clip]{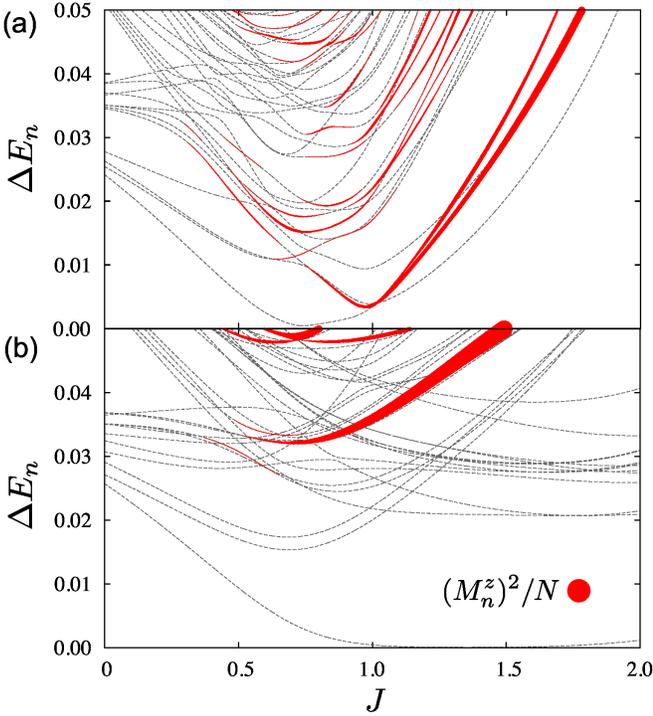}
\caption{(Color online) 
Excitation energies as functions of $J$ for (a) the effective Hamiltonian in Eq.~(\ref{eq:H_eff}) and (b) the model with $J_p'=0$ for an octamer. Matrix element $(M_n^z)^2/N$ for each excitation are also plotted by the thickness of each curve.
The parameters are chosen to be $U=10$, $\zeta=0.5$, $t'=0.3$, and $D=3$.
}
\label{fig:matrix}
\end{center}
\end{figure}

In order to elucidate the origin of the paramagnetism with a small effective moment, we here perform the spectral decomposition for the magnetic susceptibility. The $z$ component of the magnetic susceptibility in Eq.~(\ref{eq:4}) is rewritten as
\begin{align}
 \chi^{zz}=\frac{1}{N}\frac{1}{Z}\sum_{mn}\frac{e^{-\beta E_n}-e^{-\beta E_m}}{E_m-E_n}\left|\bra{\Psi_n}\sigma_{\rm total}^z\ket{\Psi_m}\right|^2,
\end{align}
where $\ket{\Psi_m}$ is the eigenstate of the Hamiltonian with the eigenvalue $E_m$, and $\sigma_{\rm total}^z=\sum_i\sigma_i^z$. Then, at low temperature, the susceptibility is approximately given as 
\begin{align}
 \chi^{zz}\sim \frac{1}{N}\sum_n\frac{1-e^{-\beta\Delta E_n}}{\Delta E_n}(M_n^z)^2,
\label{eq:chi_lim_0}
\end{align}
where $\Delta E_n=E_n-E_0$ is the excitation gap between the ground state $\ket{\Psi_0}$ and the $n$-th excited state $\ket{\Psi_n}$,
 and $M_n^z=|\bra{\Psi_n}\sigma_{\rm total}^z\ket{\Psi_0}|$ is the matrix element of $\sigma_{\rm total}^z$ between the ground state $\ket{\Psi_0}$ and the $n$-th excited state $\ket{\Psi_n}$. 
From Eq.~(\ref{eq:chi_lim_0}), if the excitation energy $\Delta E_n$ is much smaller than the temperature, the susceptibility includes a Curie-like contribution as
\begin{align}
 \chi^{zz}\sim \beta \frac{(M_n^z)^2}{N}.
\label{eq:chi_lim_1}
\end{align}
Here, the square root of $(M_n^z)^2/N$ corresponds to the effective moment.
On the other hand, when the excitation energy $\Delta E_n$ is much larger than the temperature, the susceptibility is given by
\begin{align}
 \chi^{zz}\sim \sum_n \frac{1}{\Delta E_n} \frac{(M_n^z)^2}{N}.
\label{eq:chi_lim_2}
\end{align}
This does not depend on the temperature. Hence, in the limit of $T\to 0$, the susceptibility shows Van Vleck-type behavior (or nonmagnetic behavior when $M_n^z=0$ for all $n$) if the ground state is not degenerate. Although this indicates that the slope of susceptibility in $\beta$ (effective moment) becomes zero in the limit of $T\rightarrow 0$, paramagnetic behavior is expected when the temperature is comparable or larger compared to an excitation energy for the state in which $(M_n^z)^2/N$ is nonzero, as inferred in Eq.~(\ref{eq:chi_lim_1}).

Figure~\ref{fig:matrix}(a) shows the excitation energies $\Delta E_n$ as a function of $J$. 
We also present the values of $(M_n^z)^2/N$ by the thickness of the curves in this figure. 
Although the lowest excitation energy takes minimum at $J\sim 0.8$, the matrix element $(M_n^z)^2/N$ for the first excited state is almost zero. 
On the other hand, at $J\simeq 1.0$, the matrix element $(M_n^z)^2/N$ takes a substantial value $\sim 0.1$, while the excitation energy becomes $\sim 0.003$. This result leads to the paramagnetic behavior with a small effective moment $0.3\lesssim \sqrt{0.1}$ in the temperature range of $T\gtrsim 0.003$ ($\beta \lesssim 300$), which is consistent with the temperature dependence of the susceptibility shown in Figs.~\ref{fig:sus_full}(b) and \ref{fig:sus_full}(e).

In contrast, when the $J_p'$ term is neglected, there are no such low-energy excited states which contribute to the paramagnetic behavior, as shown in Fig.~\ref{fig:matrix}(b): the lowest-energy state with a substantial value of $(M_n^z)^2/N$ appears only in the high-energy region of $\Delta E_n \gtrsim 0.03$. This result indicates that the paramagnetic behavior does not appear below $T\sim 0.03$, which is consistent with the temperature dependence of the susceptibility shown in Fig.~\ref{fig:sus_KH}.


\section{Discussion}\label{sec:discussion}

Our results provide a different picture of the octamer state in CuIr$_2$S$_4$ from the orbital Peierls scenario proposed in the previous study.~\cite{Khomskii2005} 
The previous scenario focused on the ordering of $5d$ orbitals under the tetragonal distortion while neglecting the spin-orbit coupling. 
It concluded that the octamer state is a conventional spin-singlet state composed of dimers driven by orbital ordering. 
The dimerization is caused by the Peierls-type mechanism, which is essentially the instability appearing in the weak coupling limit. 
In contrast, our theory is based on the model including both the strong electron correlation and the spin-orbit coupling. 
Our effective model is derived by the perturbation from the strong coupling limit, where these two energy scales are much larger than the transfer integrals. 
In this sense, our approach is complementary to the previous weak-coupling approach. 
In addition to the conventional spin-singlet state, which may be adiabatically connected to the orbital Peierls singlet state, our result brings about a qualitatively new state with dominant quadrupolar correlations. 
This new state is induced by the symmetric off-diagonal exchange interaction, which is enhanced by the $d$-$p$-$d$ indirect hopping, the spin-orbit coupling, and the Hund's-rule coupling.

Our analysis including the spin-orbit coupling concludes that the low-temperature behavior of the magnetic susceptibility strongly depends on the ground state. 
In the conventional spin-singlet region, the susceptibility is essentially zero at low temperature, reflecting the gap opening. 
On the other hand, in the quadrupolar region, it saturates at a nonzero value, corresponding to the Van Vleck contribution. 
The interesting observation is that the susceptibility exhibits peculiar paramagnetic behavior with a renormalized effective moment in the transient parameter region between the two regimes. 
This apparent weak paramagnetism is unusual for a finite-size cluster with strong dimerization.

In the present study, we neglect the tetragonal distortion for simplicity. The tetragonal distortion leads to additional interactions to the effective Hamiltonian in Eq.~(\ref{eq:H_eff}) through the crystal field splitting.~\cite{Yamaji2014}
Although the additional interactions slightly modulate the wave function of the singlet and quadrupolar states, the competition between these states is expected to occur under the tetragonal distortion. Therefore, we anticipate that the paramagnetism with a small effective moment emerges at low temperature even in the presence of the realistic tetragonal distortion.
Further quantitative arguments require the detailed estimates of the model parameters. We also neglected the coupling between the octamers. While the coupling is expected to be very small as it is mediated by the nonmagnetic Ir$^{3+}$ cations, it will be interesting to consider the effect of the inter-octamer coupling on the fate of the remnant paramagnetism at lower temperatures.

Recently, a $\mu$SR experiment suggested that a weak paramagnetism is persistent below $\sim 100$~K in the octamer phase.~\cite{Kojima2014} This behavior was confirmed not to originate from magnetic impurities but to be attributed to Ir moments in octamers. Our results on the weak paramagnetism may give a possible explanation of this peculiar behavior.
In our results, suppose the transfer integral $t=0.1-1$~eV, the paramagnetic behavior with a small effective moment appears below several tens K. 

Experimentally, it was deduced that the paramagnetism dominantly comes from Ir$^{4+}$ cations on the $x$ bonds in Fig.~\ref{fig:octamer} where the coordination number is three. 
The effective moment was estimated to be $0.085(3)\mu_{\rm B}$ at each site. 
In our results, the (averaged) effective moment is estimated to be, at most, 30\% 
of an isolated Ir$^{4+}$ magnetic moment and becomes smaller depending on the parameters. 
We also calculate the local magnetic susceptibility at each site of the octamer (not shown). We find that four corner sites of the octamer where the coordination number is two have a larger contribution than the sites on the $x$ bonds. This tendency is opposite to the experimental observation.

Once our theory applies, it provides a prediction which can be tested in experiments. The paramagnetic behavior is realized by the competition between the two different types of magnetic states as explained before. This competition can be controlled by the microscopic parameters, such as the transfer integrals, the spin-orbit coupling, and the Hund's-rule coupling. Particularly, the transfer integrals are sensitive to the lattice constant and structure.
We expect that the effective moment as well as the temperature dependence of the susceptibility is sensitively changed by applying pressure.


\section{Summary}\label{sec:summary}

In summary, we have studied an effective quasispin model for understanding the low-energy physics of the low-temperature octamer phase in the spinel compound CuIr$_2$S$_4$. 
We have derived the model for the $j_{\rm eff}=1/2$ states under the strong spin-orbit coupling by perturbation expansion in the strong coupling limit in terms of the transfer integrals of both $d$-$d$ direct and $d$-$p$-$d$ indirect hopping. 
The model is given in the form of an extended Kitaev-Heisenberg model, which includes an additional term of the symmetric off-diagonal interaction originating from the perturbation processes including both the direct and indirect hopping. 
We have analyzed the ground-state and finite-temperature properties of the effective model by the exact diagonalization. 

The main finding is that the model exhibits competition between the spin-singlet and quadrupolar states while changing the transfer integrals, Hund's-rule coupling, and spin-orbit coupling. 
In the spin-singlet region, the temperature dependence of the magnetic susceptibility shows gapped behavior with strong suppression at low temperature, whereas in the quadrupolar region it shows the Van Vleck-type behavior with a saturation to a nonzero value. 
Interestingly, in the competing region, we have found peculiar paramagnetic behavior with a renormalized small moment at low temperature. 
We have also clarified that the additional symmetric off-diagonal interaction plays an important role in this remnant paramagnetism.

The present results offer a different picture from the orbital Peierls scenario proposed previously.~\cite{Khomskii2005} 
Our theory is based on the strong coupling picture with the formation of local moments, while the previous scenario utilized an instability in the weak-coupling band picture. 
Although our results give no information on the stabilization mechanism of octamers, our finding of the remnant paramagnetism might explain the recent $\mu$SR experiment. 
To test our scenario, magnetic measurements in an external pressure will be interesting.

The competition between the spin-singlet and quadrupolar states will give rise to further interesting physics. 
Especially, the quantum phase included in the wave function for the quadrupolar state might lead to novel phenomena, not only in the insulating state but also in the metallic state, such as anomalous transport phenomena in the vicinity of the metal-insulator transition. 
Indeed, the octamer insulating state with charge ordering is collapsed by Zn doping to Cu sites in CuIr$_2$S$_4$, and the system becomes conductive and even exhibits the superconductivity at low temperature.~\cite{Suzuki1999,Cao2001} 
The mechanism of the superconductivity remains elusive. 
On the other hand, Se doping to S sites also makes the system conductive, but no superconductivity appears.~\cite{Nagata1998} 
It will be interesting to examine the effect of quadrupolar correlations competing with spin-singlet formation on the metal-insulator transition for understanding the mysterious properties in the doped compounds including the superconductivity.
A complementary weak-coupling approach including the spin-orbit coupling will also be interesting to clarify the mechanism of the superconductivity as well as the octamer formation.

\begin{acknowledgments}
The authors thank R.~Kadono and K.~M.~Kojima for fruitful discussion and the information on the experiments. They also thank T.~Momoi for helpful discussions. One of the authors (J.N.) acknowledges the financial support from the Japan Society for the Promotion of Science. This research was supported by KAKENHI (No. 24340076 and No. 24740221), the Strategic Programs for Innovative Research (SPIRE), MEXT, and the Computational Materials Science Initiative (CMSI), Japan.
\end{acknowledgments}


\begin{thebibliography}{99} 


\bibitem{Tokura1994}
Y. Tokura, A. Urushibara, Y. Moritomo, T. Arima, A. Asamitsu, G. Kido, and N. Furukawa,
\journal{\JPSJ}{63}{3931}{1994}.


\bibitem{Kimura2003}
T. Kimura, T. Goto, H. Shintani, K. Ishizaka, T. Arima, and Y. Tokura,
\journal{Nature}{426}{55}{2003}.


\bibitem{Kamihara2008}
Y. Kamihara, T. Watanabe, M. Hirano, and H. Hosono,
\journal{J. Am. Chem. Soc.}{130}{3296}{2008}.


\bibitem{Ueda1997}
Y. Ueda, N. Fujiwara, and H. Yasuoka,
\journal{\JPSJ}{66}{778}{1997}.

\bibitem{Mamiya1997}
H. Mamiya, M. Onoda, T. Furubayashi, J. Tang, and I. Nakatani,
\journal{J. Appl. Phys.}{81}{5289}{1997}.




\bibitem{Schmidt2004}
M. Schmidt, W. Ratcliff, P. G. Radaelli, K. Refson, N. M. Harrison, and S.-W. Cheong,
\journal{\PRL}{92}{056402}{2004}.

\bibitem{Horibe2006}
Y. Horibe, M. Shingu, K. Kurushima, H. Ishibashi, N. Ikeda, K. Kato, Y. Motome, N. Furukawa, S. Mori, and T. Katsufuji,
\journal{\PRL}{96}{086406}{2006}.



\bibitem{Nagata1994}
S. Nagata, T. Hagino, Y. Seki, and T. Bitoh,
\journal{\PSCA}{B 194-196}{1077}{1994}


\bibitem{Matsuno1997}
J. Matsuno, T. Mizokawa, A. Fujimori, D. A. Zatsepin, V. R. Galakhov, E. Z. Kurmaev, Y. Kato, and S. Nagata,
\journal{\PRB}{55}{R15979}{1997}.


\bibitem{Kumagai2000}
K. Kumagai, K. Kakuyanagi, R. Endoh, and S. Nagata,
\journal{\PSCA}{C 341-348}{741}{2000}

\bibitem{Furubayashi1994}
 T. Furubayashi, T. Matsumoto, T. Hagino, and S. Nagata,
\journal{\JPSJ}{63}{3333}{1994}.

\bibitem{Radaelli2002}
 P. G. Radaelli, Y. Horibe, M. J. Gutmann, H. Ishibashi, C. H. Chen, R. M. Ibberson, Y. Koyama, Y.-S. Hor, V. Kirykukhin, and S.-W. Cheong,
\journal{Nature}{416}{155}{2002}.

\bibitem{Wang2004}
N. L. Wang, G. H. Cao, P. Zheng, G. Li, Z. Fang, T. Xiang, H. Kitazawa, and T. Matsumoto,
\journal{\PRB}{69}{153104}{2004}.

\bibitem{Croft2003}
 M. Croft, W. Caliebe, H. Woo, T. Tyson, D. Sills, Y.-S. Hor, S.-W. Cheong, V. Kirykukhin, and S.-J. Oh,
\journal{\PRB}{67}{201102(R)}{2003}.

\bibitem{Takubo2005}
K. Takubo, S. Hirata, J.-Y. Son, J. W. Quilty, T. Mizokawa, N. Matsumoto, and S. Nagata,
\journal{\PRL}{95}{246401}{2005}.

\bibitem{Nagata1998}
S. Nagata, N. Matsumoto, Y. Kato, T. Furubayashi, T. Matsumoto, J. P. Sanchez, and P. Vulliet,
\journal{\PRB}{58}{6844}{1998}.



\bibitem{Khomskii2005}
D. I. Khomskii and T. Mizokawa,
\journal{\PRL}{94}{156402}{2005}.

\bibitem{Kim2008}
B. J. Kim, Hosub Jin, S. J. Moon, J.-Y. Kim, B.-G. Park, C. S. Leem, Jaejun Yu, T. W. Noh, C. Kim, S.-J. Oh, J.-H. Park, V. Durairaj, G. Cao, and E. Rotenberg,
\journal{\PRL}{101}{076402}{2008}.

\bibitem{Kim2009}
B. J. Kim, H. Ohsumi, T. Komesu, S. Sakai, T. Morita, H. Takagi, and T. Arima,
\journal{Science}{323}{1329}{2009}.

\bibitem{Jackeli2009}
G. Jackeli and G. Khaliullin,
\journal{\PRL}{102}{017205}{2009}.

\bibitem{Kitaev2006}
A. Kitaev,
\journal{Ann. Phys. (N.Y.)}{321}{2}{2006}.

\bibitem{Chaloupka2010}
J. Chaloupka, G. Jackeli, and G. Khaliullin,
\journal{\PRL}{105}{027204}{2010}.

\bibitem{Liu2011}
X. Liu, T. Berlijn, W.-G. Yin, W. Ku, A. Tsvelik, Y.-J. Kim, H. Gretarsson, Y. Singh, P. Gegenwart, and J. P. Hill,
\journal{\PRB}{83}{220403(R)}{2011}.

\bibitem{Jiang2011}
H.-C. Jiang, Z.-C. Gu, X.-L. Qi, and S. Trebst,
\journal{\PRB}{83}{245104}{2011}.

\bibitem{Reuther2011}
J. Reuther, R. Thomale, and S. Trebst,
\journal{\PRB}{84}{100406}{2011}.

\bibitem{Choi2012}
S. K. Choi, R. Coldea, A. N. Kolmogorov, T. Lancaster, I. I. Mazin, S. J. Blundell, P. G. Radaelli, Y. Singh, P. Gegenwart, K. R. Choi, S.-W. Cheong, P. J. Baker, C. Stock, and J. Taylor,
\journal{\PRL}{108}{127204}{2012}.

\bibitem{Ye2012}
F. Ye, S. Chi, H. Cao, B. C. Chakoumakos, J. A. Fernandez-Baca, R. Custelcean, T. F. Qi, O. B. Korneta, and G. Cao,
\journal{\PRB}{85}{180403(R)}{2012}.



\bibitem{Rau2013}
J. G. Rau, E. K.-H. Lee, and H.-Y. Kee,
\journal{\PRL}{112}{077204}{2014}.


\bibitem{Kojima2014}
K. M. Kojima, R. Kadono, M. Miyazaki, M. Hiraishi, I. Yamauchi, A. Koda, Y. Tsuchiya, H. S. Suzuki, H. Kitazawa, 
\journal{\PRL}{112}{087203}{2014}.



\bibitem{Shannon2006}
N. Shannon, T. Momoi, and P. Sindzingre,
\journal{\PRL}{96}{027213}{2006}.


\bibitem{Yamaji2014}
Y. Yamaji, Y. Nomura, M. Kurita, R. Arita, and M. Imada,
arXiv:1402.1030.

\bibitem{Foyevtsova2013}
K. Foyevtsova, H. O. Jeschke, I. I. Mazin, D. I. Khomskii, and R. Valenti,
\journal{\PRB}{88}{035107}{2013}.


\bibitem{Suzuki1999}
H. Suzuki, T. Furubayashi, G. Cao, H. Kitazawa, A. Kamimura, K. Hirata, and T. Matsumoto, 
\journal{\JPSJ}{68}{2495}{1999}.

\bibitem{Cao2001}
G. Cao, T. Furubayashi, H. Suzuki, H. Kitazawa, T. Matsumoto, and Y. Uwatoko,
\journal{\PRB}{64}{214514}{2001}.


\end{thebibliography}


\end{document}